\documentclass[12pt]{iopart}

\usepackage{epstopdf}
\usepackage{graphicx, setspace}
\usepackage[colorlinks=true,linkcolor=blue]{hyperref}
\begin{document}

\title[]{Phase Mixing of Relativistically Intense Longitudinal Wave Packets in a Cold Plasma}

\author{Arghya Mukherjee and Sudip Sengupta}

\address{Institute for Plasma Research, Bhat, Gandhinagar, 382428, India}
\ead{arghya@ipr.res.in}
\vspace{10pt}

\begin{abstract}
Phase mixing of relativistically intense longitudinal wave packets in a cold homogeneous unmagnetized plasma has been studied analytically and numerically using Dawson Sheet Model. A general expression for phase mixing time ($\omega_pt_{mix}$) as a function of amplitude of the wave packet($\delta$) and width of the spectrum($\Delta k/k$) has been derived. It is found that phase mixing time crucially depends on the relative magnitude of amplitude ``$\delta$" and the spectral width ``$\Delta k/k$". For $\Delta k/k \leq 2\omega_p^2\delta ^2/c^2k^2$, $\omega_pt_{mix}$ scales with $\delta$ as $\sim 1/\delta ^5$, whereas for $\Delta k/k > 2\omega_p^2\delta ^2/c^2k^2$, $\omega_pt_{mix}$ scales with $\delta$ as $\sim 1/\delta ^3$, where $\omega_p$ is the non-relativistic plasma frequency and $c$ is the speed of light in vacuum. We have also verified the above theoretical scalings using numerical simulations based on Dawson Sheet Model.
\end{abstract}

%
%
%
\maketitle
%
%
\section{Introduction}
The study of relativistically intense longitudinal plasma waves and their space-time evolution is an area of intense research in plasma physics because of its application to a broad range of physical problems related to laser plasma interaction\cite{ref1,ref2,ref3,ref4,ref5,ref6,ref7,ref8,ref9,a3,a4}, astrophysical plasmas\cite{ref10,ref11} and inertial confinement thermonuclear fusion\cite{a1,a2}. For example study of nonlinear plasma waves is important from the point of view of wakefield acceleration where the wake wave is excited either by passing laser pulses or bunches of relativistic electron beams through a plasma chamber\cite{ref1,ref9,ref14,ref15,ref16}. Amplitude of these relativistically intense space charge waves is limited by the phenomenon of wave breaking, which occurs via a well known process called phase mixing \cite{ref17,ref18,ref19,ref20}. Phase mixing results in crossing of neighbouring electron orbits which is caused by temporal dependence of phase difference between oscillating electrons constituting the oscillation/wave\cite{ref18}. This temporal dependence of phase difference between neighbouring oscillating electrons arises because of background density inhomogeneities (either fixed\cite{ref21,ref22} or self-generated\cite{ref23}) and/or because of relativistic mass variation effects\cite{ref17,ref19,ref20,ref24}. The process of phase mixing leading to wave breaking not only limits the maximum achievable electric field in laser/beam driven wakefield experiments, but also is applicable to some electron injection schemes\cite{ref25,ref26,ref27}, where the wake wave moves along a density gradient and traps electrons by breaking. Further, in relation to, laser/beam driven wakefield experiments it has been recently pointed out\cite{ref19} that if the phase mixing time (wave breaking time) is shorter than the dephasing time of an electron in the wake wave, then maximum energy gain cannot be achieved as the wake wave gets damped because of phase mixing before the dephasing time is reached. Therefore a thorough understanding of the phase mixing process and estimation of phase mixing time (wave breaking time) of relativistically intense plasma waves is relevant for these experiments.\\
In a typical laser/beam plasma interaction experiment, a spectrum of relativistically intense plasma waves with an arbitrary spread in $\Delta k$(and hence in $v_{ph}$) is excited because of group velocity dispersion and nonlinear distortion of light pulse near the critical layer\cite{ref17,ref18}. Such a wave packet exhibits phase mixing and breaks at arbitrarily small amplitudes\cite{ref17}. By studying the space-time evolution of two relativistically intense waves having wave numbers separated by an amount $\Delta k$, authors in ref.\cite{ref17} showed that, in general a wave packet having amplitude $\delta$ and spectral width $\Delta k$ will phase mix and break in a time scale given by $\omega_pt_{mix} \sim \left\lbrace \frac{3}{64}(3\omega_p ^2 \delta ^3/c^2k^2)[|\Delta k/k|/(|1 + \Delta k /k|)](1 + 1/|1 + \Delta k/k|) \right\rbrace ^{-1}$. This expression shows that in the limit $\Delta k/k \rightarrow 0$, $\omega_pt_{mix} \rightarrow \infty$ i.e a sinusoidal wave will not undergo phase mixing. This is contrary to present understanding that relativistically intense plasma waves in a cold homogeneous plasma with immobile ions always phase mixes and breaks\cite{ref17,ref24} at an arbitrary low amplitude; except for the singular case of longitudinal Akhiezer-Polovin mode whose amplitude is limited to $\frac{eE_{WB}}{m\omega_{pe}c} \sim \sqrt{2}(\gamma_{ph} -1)^{1/2}$. Here, $\gamma_{ph} = 1/\sqrt{(1 - v_{ph}^2/c^2)}$ is the Lorentz factor associated with the phase velocity $v_{ph}$, $e$ and $m$ are, respectively, the charge and mass of an electron. A longitudinal Akhiezer - Polovin mode is a very special combination of $\omega$,$k$ ($\omega$ = Frequency, $k$ = Wavenumber) and its harmonics such that they propagate together as a coherent nonlinear structure with a constant phase velocity\cite{ref28}. Even a longitudinal Akhiezer-polovin mode 
phase mixes and breaks at an arbitrarily low amplitude when perturbed longitudinally\cite{ref19,ref20}.\\
In order to resolve this anomaly, in the present paper we extend the calculation of ref.\cite{ref17} and show both analytically and numerically that phase mixing occurs even in the limit $\Delta k/k \rightarrow 0$. In fact, we clearly delineate regimes where the phase mixing formula presented in ref.\cite{ref17} holds.\\
In ref.\cite{ref17} , the relativistic equation of motion of an electron is derived using Dawson sheet model\cite{ref29,ref30} and solved in the weakly relativistic limit using Krylov and Bogoliubov method of averaging\cite{ref31}. The relativistically correct frequency of oscillation is obtained upto second order in wave amplitude ``$\delta$". The expression for frequency thus derived, because of relativistic mass variation effects, clearly exhibits spatial dependency, which is a signature of phase mixing. As noted in the previous paragraph, the spatial dependency vanishes in the limit $\Delta k/k \rightarrow 0$, indicating that a sinusoidal wave does not phase mix. To resolve this, in section \ref{section2} of this paper, we extend the calculation of ref.\cite{ref17} and compute the frequency correct upto fourth order in oscillation amplitude ``$\delta$" using Lindstedt - Poincar\'{e} method\cite{ref32}. This improved calculation of frequency exhibits spatial dependency even in the limit of $\Delta k/k \rightarrow 0$. Using this modified frequency and using Dawson's argument\cite{ref21}, phase mixing time is estimated, both in small and large $\Delta k/k$ limit. In section \ref{section3}, we verify the analytically derived scalings of phase mixing time on amplitude ``$\delta$", using numerical simulations based on Dawson sheet model\cite{ref29,ref30}. Finally in section \ref{section4}, we present a summary of our work.

\section{Equation of Motion and Its Solution} \label{section2}
According to the Dawson sheet model description of a cold plasma\cite{ref29,ref30}, electrons are assumed to be infinite sheets of charges embedded in a cold immobile positive ion background. Evolution of any coherent mode can be studied in terms of oscillating motion of these sheets about their equilibrium positions. Let $x_{eq}$ and $\xi(x_{eq},t)$, respectively, be the equilibrium position and displacement from the equilibrium position of an electron sheet. In terms of $x_{eq}$ and $\xi(x_{eq},t)$, the associated fluid quantities, viz., number density, velocity and self consistent electric field can, respectively, be written as $n(x_{eq},t) = n_0/(1 + \partial \xi/\partial x_{eq})$, $v(x_{eq},t) = \dot{\xi}$ and $E(x_{eq},t) = 4\pi en_0\xi$, where $n_0$ is the equilibrium density of electrons. Here, dot represents differentiation w.r.t time $t$. The Euler coordinate of the electron sheet is given by $x(x_{eq},t) = x_{eq} + \xi(x_{eq},t)$. So, once $\xi(x_{eq},t)$ is found, the space time evolution of a mode is solved in principle. $\xi(x_{eq},t)$ can be obtained by solving the relativistic equation of motion of a sheet which can be written as\cite{ref17}
\begin{equation}
\frac{\ddot{\xi}}{(1 - \dot{\xi} ^2)^{3/2}} + \xi = 0  \label{eq1}
\end{equation}
Here, we have used the following normalization: $t \rightarrow \omega_{pe}t$, $\xi \rightarrow \frac{\omega_{pe}\xi}{c}$, $\dot{\xi} \rightarrow \frac{\dot \xi}{c}$, $E \rightarrow \frac{eE}{m\omega_{pe}c}$. In weakly relativistic limit, Eq.\ref{eq1} transforms to\cite{ref17}
\begin{equation}
\ddot{\xi} + \xi - \frac{3}{2}\xi \dot{\xi} ^2 \approx 0   \label{eq2}
\end{equation}
Eq.\ref{eq2} is solved using Lindstedt - Poincar\'{e} Preturbation Technique\cite{ref32} by expanding in series the displacement $\xi$ and the oscillation frequency $\Omega$: $\xi = \xi _0 + \xi _1 + \xi _2 + ...$ and $\Omega ^2 = 1 + \omega _1^2 + \omega _2 ^2 + ...$\cite{ref32} . Initial conditions are taken same as in Ref. \cite{ref17}
\numparts
\begin{equation}
n_e(x,0) = n_0\left[1 + \delta cos\left(\frac{\Delta k}{2}x\right)cos\left(k+\frac{\Delta k}{2}\right)x\right]  \label{eq3a}
\end{equation}
and
\begin{equation}
v_e(x,0) = \frac{\omega_{pe}\delta}{2}\left[\frac{1}{k}cos(kx) + \frac{1}{k + \Delta k}cos(k+\Delta k)x \right]     \label{eq3b}
\end{equation}
\endnumparts
 The zeroth order solution to Eq.\ref{eq2} is $\xi _0 = \xi (x_{eq})cos[\Omega t + \phi_0 (x_{eq})]$ with $\xi (x_{eq})$ being the oscillation amplitude. $\xi(x_{eq})$ and $\phi_0(x_{eq})$ are respectively given by
  \numparts
\begin{equation}
\xi(x_{eq}) = \frac{\omega_p\delta}{2ck}\left[1 + \frac{k^2}{(k + \Delta k)^2} + \frac{2k}{(k + \Delta k)}cos(\Delta kx_l)\right]^{1/2}  \label{eq4a}
\end{equation}
and
\begin{equation}
\phi_0(x_{eq}) = tan^{-1}\left\lbrace \frac{\left[ \frac{cos(kx_l)}{k} + \frac{cos(k+\Delta k)x_l}{(k + \Delta k)} \right]}{\left[ \frac{sin(kx_l)}{k} + \frac{sin(k+\Delta k)x_l}{(k + \Delta k)} \right]}\right\rbrace     \label{eq4b}
\end{equation}
\endnumparts
where $x_l = x_{eq} + \xi(x_{eq},0)$, the initial position of a sheet.\\
The solution to Eq.\ref{eq2} correct upto $\xi_1$ can be written as
\begin{eqnarray}
\xi(x_{eq},t) = \xi(x_{eq})cos(\Omega t + \phi_0) + \frac{3\xi(x_{eq})^3}{16}cos(\Omega t + 2\phi_0) \nonumber \\- \frac{\xi(x_{eq})^3}{16}cos(\Omega t - 2\phi_0)  
- \frac{\xi(x_{eq})^3}{8}cos{2(\Omega t + \phi_0)} \label{eq5}
\end{eqnarray}
 The frequency $\Omega$ is determined from the condition that there are no secular resonant terms in the equations for $\xi _1$, $\xi _2$... We find that the oscillation frequency correct upto the fourth order of the oscillation amplitude $\delta$ is given by
\begin{eqnarray}
\Omega \approx 1 - \frac{3\omega_p ^2\delta ^2}{64c^2}\left[\frac{1}{k^2} + \frac{1}{(k+\Delta k)^2} + \frac{2cos(\Delta kx_l)}{k(k+\Delta k)}\right] - \frac{3\omega_p ^4\delta ^4}{1024c^4} \times \nonumber \\ \left[3\left\lbrace \frac{cos(kx_l)}{k} + \frac{cos(k + \Delta k)x_l}{(k+\Delta k)} \right\rbrace^4 - \left\lbrace \frac{1}{k^2} + \frac{1}{(k+\Delta k)^2} + \frac{2cos(\Delta kx_l)}{k(k+\Delta k)}\right\rbrace ^2\right] \label{eq6}
\end{eqnarray}
The expression for frequency clearly shows spatial dependency (dependence on initial position of the sheet) for arbitrary values of $\Delta k/k$. As $\Delta k/k \rightarrow 0$, the first correction term becomes independent of sheet positions whereas the second correction term ($``\delta ^4"$ term) still retains its spatial dependence. Because of this space dependence different ``pieces" of the wave slowly go out of phase as time progresses, resulting in the phenomenon of phase mixing. Following Dawson's argument\cite{ref21}, the phase mixing time is given by
\begin{equation}
t_{mix} \sim \frac{\pi}{2}\frac{1}{\xi_{max}\frac{d\Omega}{dx_{eq}}} \label{eq7}
\end{equation}
where $d\Omega/dx_{eq}$(calculated from Eq.\ref{eq6}) is given by
\begin{eqnarray}
\frac{d\Omega}{dx_{eq}} = \frac{3\omega_p^2\delta ^2}{64c^2}\left\lbrace\frac{2sin(\Delta kx_l)\Delta k}{k(k + \Delta k)}\right\rbrace + \frac{36\omega_p^4\delta ^4}{1024c^4}\left\lbrace \frac{cos(kx_l)}{k} + \frac{cos(k+\Delta k)x_l}{(k+\Delta k)}\right\rbrace^3 \nonumber \\
\times \left\lbrace sin(kx_l) + sin(k+\Delta k)x_l\right\rbrace   \label{eq8}
\end{eqnarray}
In the above expression the small term of order $(\delta ^2 \Delta k/k)^2$ is neglected. Taking $\xi_{max} = \frac{\omega_{pe}\delta}{2c}\left[1/k + 1/(k+\Delta k)\right]$ (from Eq.\ref{eq4a}), calculating the maximum value of $\frac{d\Omega}{dx_{eq}}$ and putting them in Eq.\ref{eq7}, the final expression for phase mixing time ($t_{mix}$) stands as
\begin{eqnarray}
t_{mix} = \frac{\pi}{2}\left[\frac{3\omega_{pe}^2\delta ^3}{64c^2k^2}\left\lbrace 1 + \frac{1}{(1 + \Delta k/k)}\right\rbrace\right]^{-1}
\left[\frac{\Delta k/k}{1 + \Delta k/k} + \frac{9\sqrt{3}\omega_{pe}^2\delta ^2}{8c^2k^2}\right]^{-1} \label{eq9}
\end{eqnarray}
It is clear from above expression, that for $\Delta k/k > 2\omega_p^2\delta ^2/c^2k^2$, upto leading order, the phase mixing time is given by
\begin{equation}
t_{mix} = \frac{\pi}{2}\left[\frac{3\omega_{pe}^2\delta ^3}{64c^2k^2}\left\lbrace 1 + \frac{1}{(1 + \Delta k/k)}\right\rbrace \left\lbrace \frac{\Delta k/k}{1 + \Delta k/k} \right\rbrace \right]^{-1}   \label{eq10}
\end{equation}
This is as same expression as Eq.19 of ref.\cite{ref17}. This shows for $\Delta k/k > 2\omega_p^2\delta ^2/c^2k^2$, $t_{mix}$ scales with $``\delta"$ as $1/\delta ^3$. For $\Delta k/k \leq 2\omega_p^2\delta ^2/c^2k^2$, upto leading order, the phase mixing time is given by
\begin{equation}
t_{mix} = \frac{\pi}{2}\left[\frac{27\sqrt{3}\omega_{pe}^4\delta ^5}{512c^4k^4}\left\lbrace 1 + \frac{1}{(1 + \Delta k/k)}\right\rbrace\right]^{-1} \label{eq11}
\end{equation}
which shows that for sharply peaked wave packets, the phase mixing time $t_{mix}$ scales with $``\delta"$ as $1/\delta ^5$. In the next section, we verify these scalings using a code based on Dawson sheet model.
\section{Numerical Verification} \label{section3}
\begin{figure}[htbp]
\centering
\includegraphics[scale=0.8]{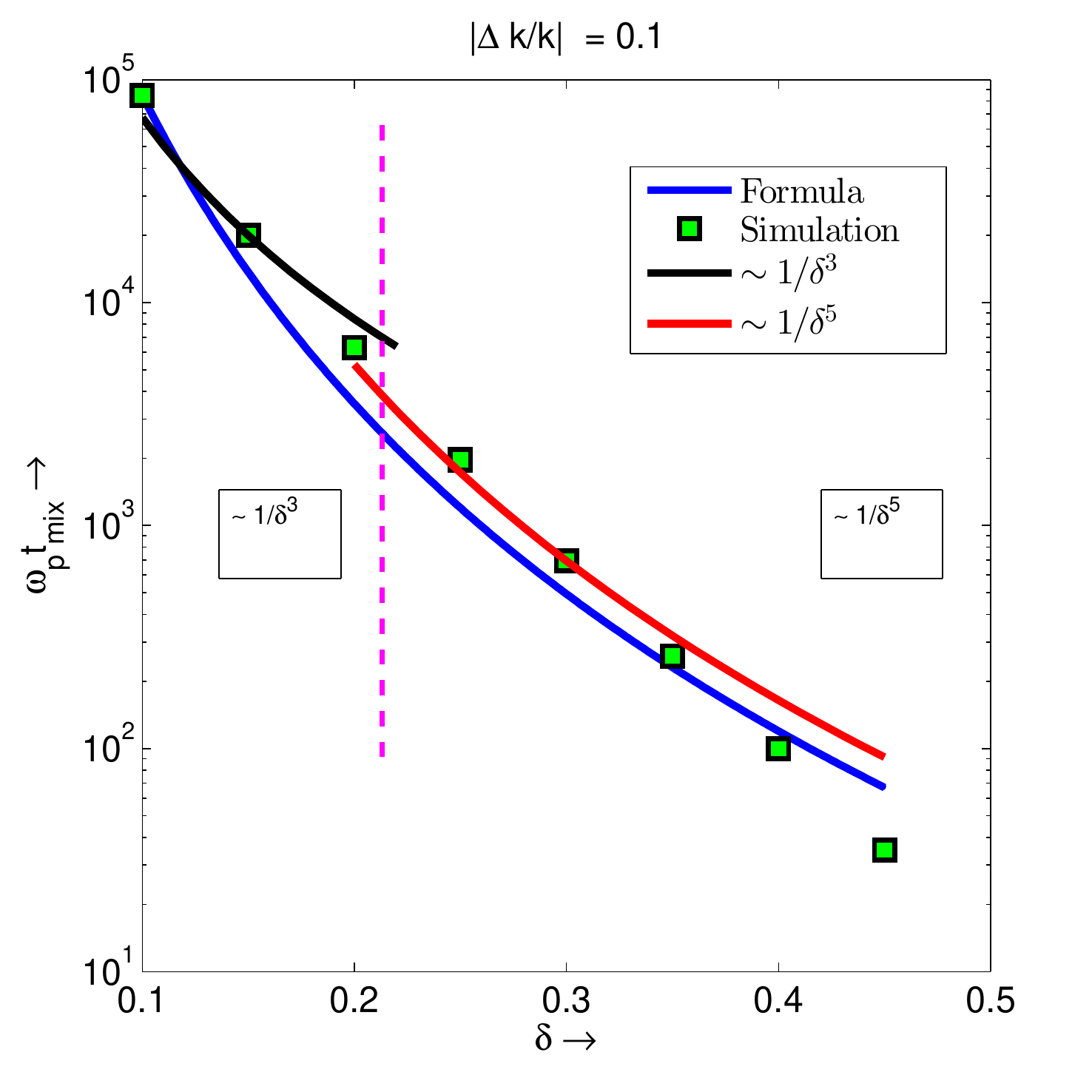}
\caption{Phase Mixing Time ($\omega_pt_{mix}$) as a function of amplitude $\delta$ for $|\Delta k/k| = 0.1$.}
\label{fig2}
\end{figure}

\begin{figure}[htbp]
\centering
\includegraphics[scale=0.8]{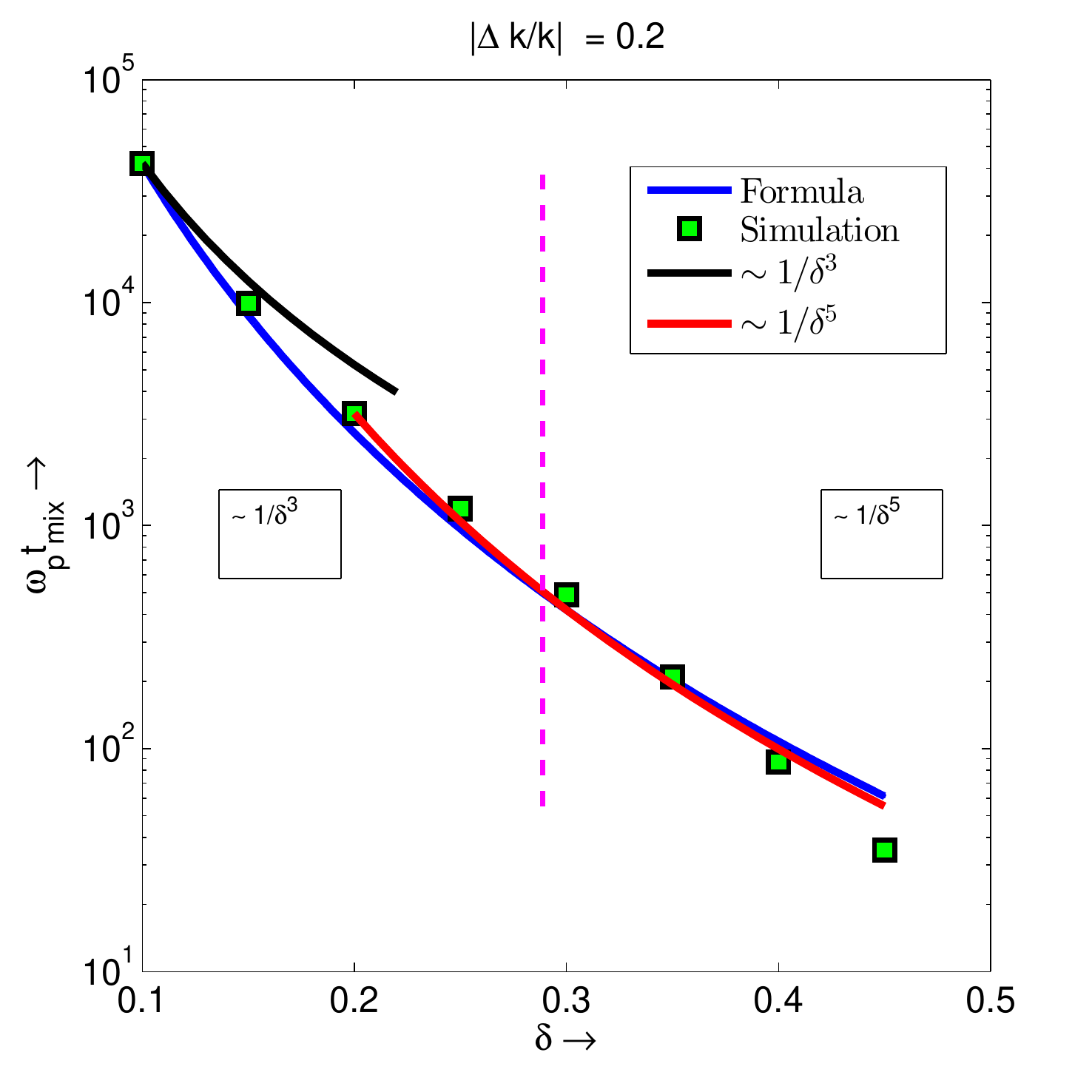}
\caption{Phase Mixing Time ($\omega_pt_{mix}$) as a function of amplitude $\delta$ for $|\Delta k/k| = 0.2$.}
\label{fig3}
\end{figure}

\begin{figure}[htbp]
\centering
\includegraphics[scale=0.8]{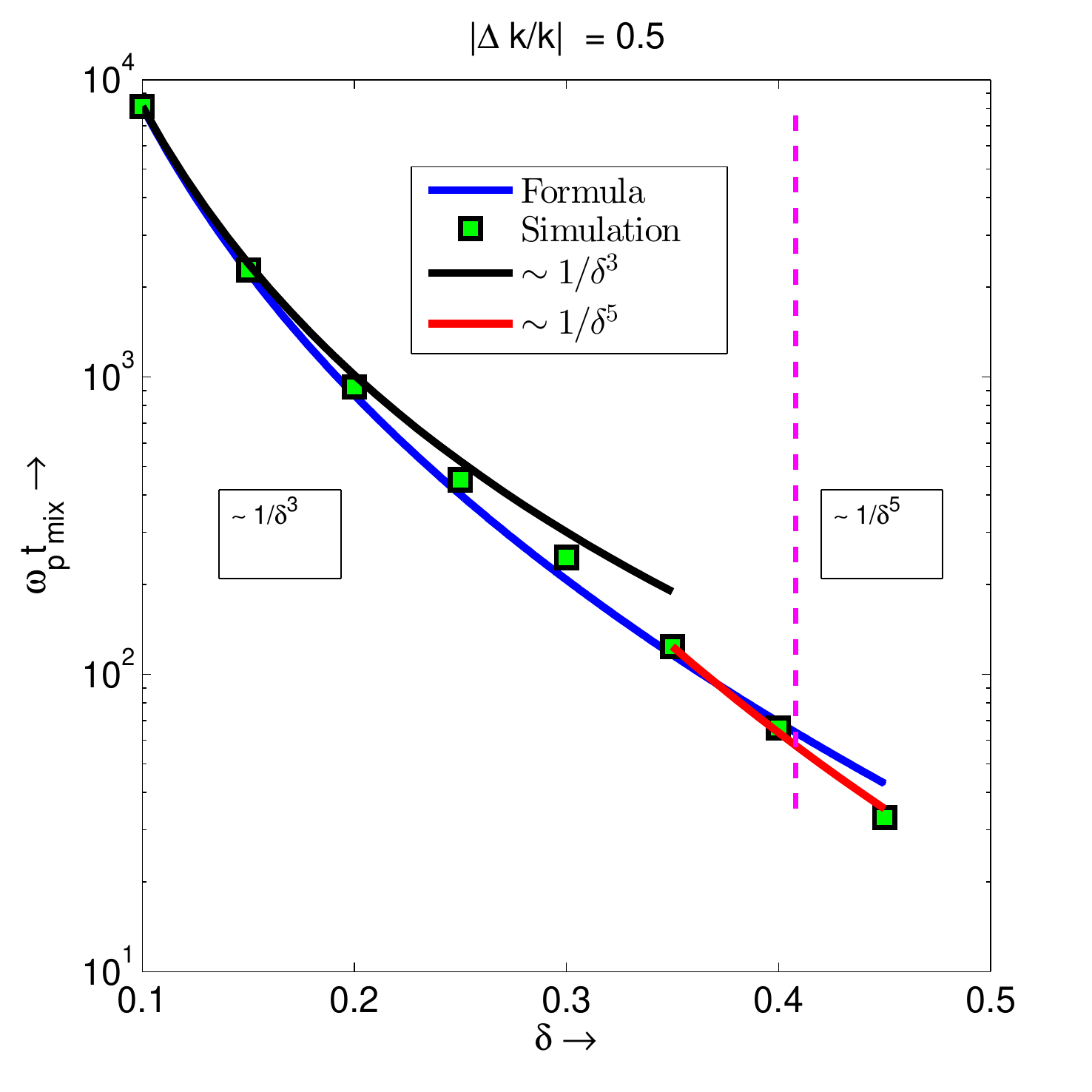}
\caption{Phase Mixing Time ($\omega_pt_{mix}$) as a function of amplitude $\delta$ for $|\Delta k/k| = 0.5$.}
\label{fig4}
\end{figure}

\begin{figure}[htbp]
\centering
\includegraphics[scale=0.8]{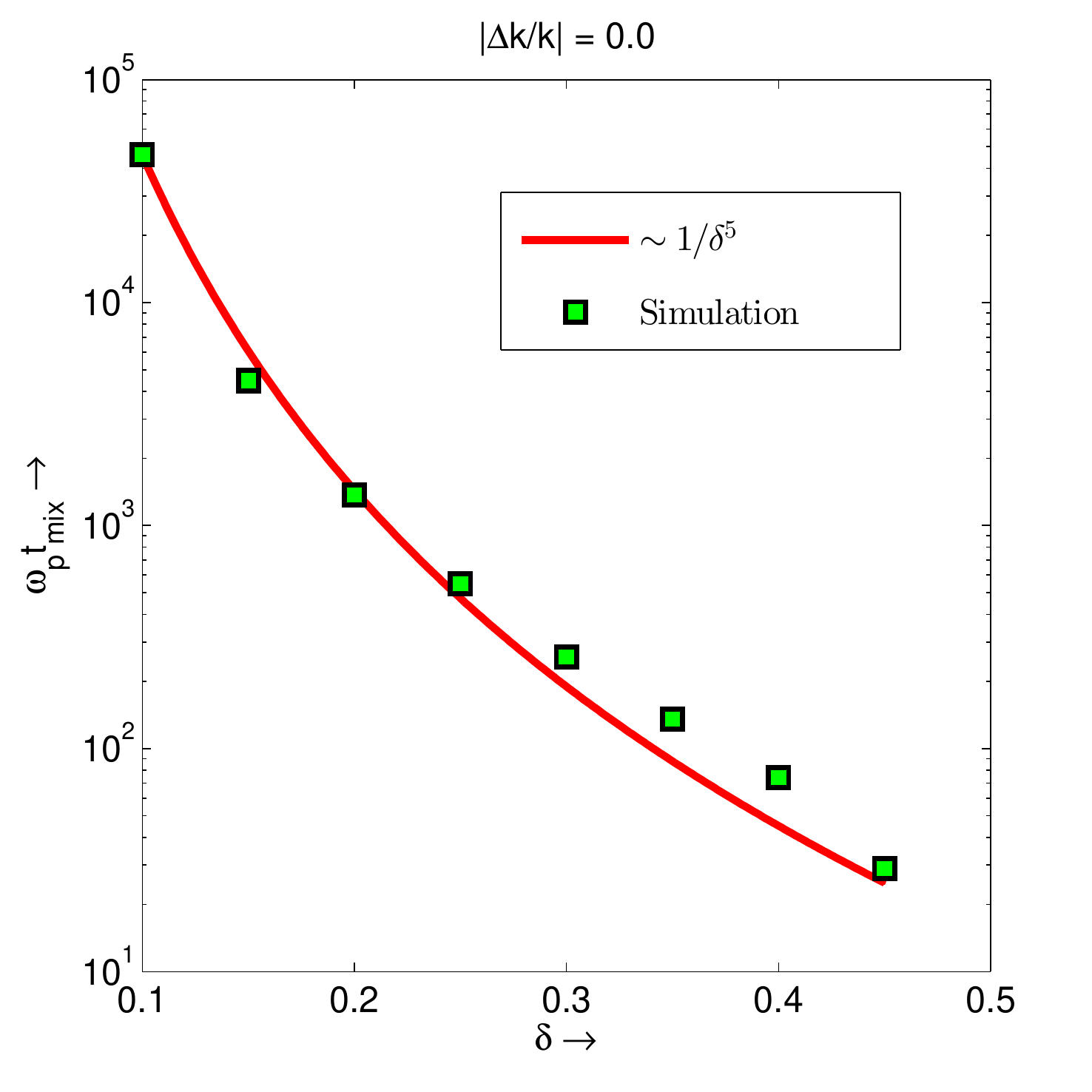}
\caption{Phase Mixing Time ($\omega_pt_{mix}$) as a function of amplitude $\delta$ for $|\Delta k/k| = 0$.}
\label{fig5}
\end{figure}

Using a code based on Dawson sheet model, we numerically verify the process of phase mixing of a longitudinal wave packet - described by it's two parameters, amplitude $\delta$ and spectral width $\Delta k/k$. For this purpose we have used a one - dimensional (1D) sheet code based on Dawson Sheet Model of a 1D plasma. In this code we have followed the motion of an array of $\sim$ 10000 electron sheets. Using initial conditions given by Eqs.\ref{eq3a}, \ref{eq3b} and using periodic boundary conditions, the equation of motion for each sheet is then solved using fourth order Runge-Kutta scheme. At each time step, ordering of sheets is checked for sheet crossing(electron trajectory crossing). Phase mixing time is measured as the time taken by any two adjacent sheets to cross over. We terminate our code at this time because the expression for {self-consistent} electric field ($E = 4\pi en_0\xi$) used in equation of motion becomes invalid beyond this point\cite{ref17,ref18,ref29,ref30}.\\
Figs.\ref{fig2} - \ref{fig4} respectively show the dependence of phase mixing time with amplitude $\delta$ for three different values of $|\Delta k/k|$; $|\Delta k/k|$ =  0.1, 0.2 and 0.5. All the figures points represent the simulation results and the blue line is the complete formula given by Eq.\ref{eq9}. Compare figure \ref{fig4} with figure 10 in ref\cite{ref17}. Note that the improved formula (Eq.\ref{eq9}) shows a much better fit to the simulation results. The vertical dashed magenta line separates the two regimes viz. $\Delta k/k > 2\omega_p^2\delta ^2/c^2k^2$ and $\Delta k/k < 2\omega_p^2\delta ^2/c^2k^2$. In the regime $\Delta k/k > 2\omega_p^2\delta ^2/c^2k^2$, the dependence of phase mixing time on $\delta$ is predominantly $\sim 1/\delta ^3$ (black line) whereas the regime $\Delta k/k \leq 2\omega_p^2\delta ^2/c^2k^2$, the dependence of phase mixing time on $\delta$ is predominantly $\sim 1/\delta ^5$ (red line). We observe, that as $|\Delta k/k|$ increases, the vertical dashed line shifts towards the right as expected from theoretical analysis. For sinusoidal case ($|\Delta k/k| = 0$), the phase mixing time scale is given by $t_{mix} = \frac{\pi}{2}\left[\frac{27\sqrt{3}\omega_{pe}^4\delta ^5}{256c^4k^4}\right]^{-1}$, obtained by putting $|\Delta k/k| = 0$ in Eq.\ref{eq11}. In Fig.\ref{fig5} we have shown the variation of phase mixing time scale as a function of $\delta$ for $|\Delta k/k| = 0$. Here the points are simulation results and the continuous line represents the scaling obtained from Eq.\ref{eq11}. In this case phase mixing time scale is always proportional to $1/\delta ^5$. In all the cases, the analytical expressions presented by Eqs.\ref{eq9} and \ref{eq11} are showing a good fit to the observed numerical results, thus vindicating our analytical results. 
\section{Summary} \label{section4}
The breaking of relativistically intense longitudinal wave packets in a cold plasma has been studied. It is shown that the phase mixing time scale $t_{mix}$ crucially depends on the relative magnitude of the amplitude of the wave packet $\delta$ and dimensionless spectral width of the wave packet $|\Delta k/k|$. For sharply peaked wave packets i.e for $\Delta k/k \leq 2\omega_p^2\delta ^2/c^2k^2$, $t_{mix}$ scales with $\delta$ as $1/\delta ^5$. For broader wave packet i.e for $\Delta k/k > 2\omega_p^2\delta ^2/c^2k^2$, $t_{mix}$ scales with $\delta$ as $1/\delta ^3$.
\section*{Acknowledgement}
We would like to acknowledge Predhiman K Kaw for useful suggestions.

\newpage
\newpage
\newpage
\newpage
\end{document}